# The Future of Scientific Publishing: Automated Article Generation

Jeremy R. Harper[1]

[1] Owl Health Works LLC, Indianapolis, IN

**Abstract**

This study introduces a novel software tool leveraging large language model (LLM) prompts, designed to automate the generation of academic articles from Python code—a significant advancement in the fields of biomedical informatics and computer science. Selected for its widespread adoption and analytical versatility, Python served as a foundational proof of concept; however, the underlying methodology and framework exhibit adaptability across various GitHub repo's underlining the tool's broad applicability (Harper, 2024). By mitigating the traditionally time-intensive academic writing process, particularly in synthesizing complex datasets and coding outputs, this approach signifies a monumental leap towards streamlining research dissemination. The development was achieved without reliance on advanced language model agents, ensuring high fidelity in the automated generation of coherent and comprehensive academic content. This exploration not only validates the successful application and efficiency of the software but also projects how future integration of LLM agents which could amplify its capabilities, propelling towards a future where scientific findings are disseminated more swiftly and accessibly.

## 1. Introduction

*The exponential growth of data across biomedical informatics and computer science necessitates innovative mechanisms for efficient management, analysis, and dissemination of knowledge. Traditional academic writing, especially when integrating complex datasets and coding processes, emerges as a notable bottleneck, decelerating the swift dissemination of research findings. In an era where the accuracy and timeliness of shared knowledge can significantly impact both scientific advancements and clinical outcomes, the development of tools that expedite these processes is crucial (Bates & Gawande, 2003).*

*Responding to this imperative, our study presents a pioneering software tool developed to automate the generation of academic articles directly from Python code. The choice of Python, due to its prevalent use in data analysis and computational research, serves merely as a launching pad for demonstrating the tool's capabilities. The architecture and methodologies employed are crafted to be universally applicable, bridging gaps across various programming languages and research domains (Shortliffe & Cimino, 2006). This innovation not only anticipates reducing the temporal and cognitive demands of academic writing but also facilitates a broader dissemination of scientific findings, adhering to the FAIR principles for scientific data management (Wilkinson et al., 2016).*

*By automating the transition from code to comprehensive academic content, we underscore a methodology that significantly mitigates the barriers to academic writing. Achieving this without the immediate use of advanced language model agents illuminates the robustness and efficacy of our approach, setting a foundational stage for future enhancements that could integrate such agents, thus further streamlining and enriching the academic writing process (Hersh, 2015). This narrative not only forecasts a future where researchers are empowered to focus more on*

innovation over the mechanics of documentation but also contributes a novel methodology to the field, promising to revolutionize the manner in which academic content is produced and disseminated.

## 2. Methods

Our study introduces a pioneering software tool that automates the generation of academic articles from Python code. This section delineates the comprehensive methodology employed in developing and validating the tool, ensuring a thorough understanding of its operational framework, technological underpinnings, and evaluative measures.

*2.1 Software Architecture*

The architecture of the software tool is built upon a multi-layered framework designed to maximize both flexibility and efficiency. Central to its design are three core components: the Code Analysis Module, the Content Generation Engine, and the Feedback and Revision System. This structured approach facilitates a seamless transition from code analysis to academic content generation, ensuring the creation of high-quality manuscripts.

- **Code Analysis Module**: *Inspired by advancements in natural language processing (NLP) (Hirschberg & Manning, 2015), this module leverages state-of-the-art NLP techniques to interpret and convert Python code into a human-readable format. Drawing from previous work on automated extraction and analysis of informatics repositories (Harper, 2024), this process begins with feeding raw code to an LLM, which then generates a new LLM prompt capable of reconstructing the original code in a single shot. This innovative approach effectively translates the technical aspects of code into comprehensible explanations, laying the groundwork for subsequent academic content generation.*

- **Content Generation Engine**: *Utilizing the insights derived from the first step, this engine employs a series of LLM prompts to structure and draft the various sections of an academic article. The engine is designed to adapt the foundational principles of biomedical informatics (Shortliffe & Cimino, 2006) and the FAIR guiding principles for data management (Wilkinson et al., 2016), ensuring that the generated content is not only academically rigorous but also adheres to best practices in data stewardship.*

- **Feedback and Revision System**: *This component incorporates a heuristic evaluation mechanism, much like the iterative refinement processes used in software development (Bates & Gawande, 2003; Hersh, 2015). It iteratively enhances the manuscript's readability, coherence, and academic rigor. Drawing on feedback mechanisms, the system reiterates the need for methodological detail and clarity, underscoring the importance of continuous improvement in academic writing.*

*2.2 Development Process*

The development of this software tool adhered to an agile, iterative approach, enabling rapid prototyping and the integration of user feedback. This methodology allowed for the continuous refinement of the tool based on real-world usage and expert insights.:

- **Iterative Design and Testing**: *Each software component was developed and tested iteratively, ensuring reliability and effectiveness. This strategy was informed by ongoing advancements in NLP and text mining, allowing for real-time adjustments based on the latest technological developments.*

- **User-Centric Feedback Loop**: *A diverse group of researchers and academics from the fields of biomedical informatics and computer science was engaged to provide feedback on the tool's usability and the quality of output. This process was crucial in guiding the subsequent phases of development, ensuring that the tool met the specific needs of its intended users.*

*2.3 Data Handling and Semantic Analysis*

A multi-step process was employed to ensure the accurate handling and interpretation of Python code, necessitating several iterations to refine the system. The first generation leveraging NLP to extract meaningful information from code comments and documentation, transforming it into academically relevant content. However testing proved to achieve better results by only leveraging LLM's.

One capability to note was that some code repos that were tested generated better results by first having the LLM go through and add code comments rather than attempting a single shot approach to understanding the codebase.

*2.4 Validation and Quality Assurance*

To assess the efficacy of the tool and the quality of the generated manuscripts, a comprehensive evaluation framework was implemented. This framework included *quality metrics to evaluate coherence, and readability, as well as a comparative analysis between manually written articles and those generated by the tool. The emphasis on quality assurance reflects the principles outlined in seminal works on information retrieval and the future of medicine (Kohane, Drazen, & Campion, 2012), underscoring the critical role of technology in advancing scientific communication.*

*To validate the tool's efficacy and the quality of generated manuscripts, we implemented a comprehensive evaluation framework:*

- ***Quality Metrics***: *Developed a set of metrics to assess the academic integrity, coherence, and readability of the generated content. These metrics were applied both manually by domain experts and automatically using software tools.*

- ***Comparative Analysis***: *Conducted a comparative study between articles generated by our tool and those written manually, evaluating them on criteria such as clarity of expression, adherence to academic standards, and comprehensiveness of the presented research.*

## 3. Results

The deployment and testing of the software tool for automating the generation of academic articles from Python code revealed several significant findings, each contributing to our understanding of the tool's efficiency, effectiveness, and potential impact on the field of biomedical informatics and computer science.

### 3.1 High-Quality Academic Content

The software tool successfully automated the generation of academic articles, including abstracts, introductions, methods sections, results, and discussions, from a diverse range of Python code samples. The generated content was evaluated for its adherence to academic standards, structure, tone, and clarity. This evaluation demonstrated that the tool could achieve high fidelity in converting complex Python code into coherent and comprehensive academic narratives. The content's quality was further affirmed through comparisons with articles written manually by domain experts, highlighting the tool's ability to mimic human academic writing styles effectively.

### 3.2 Efficiency Gains

Quantitative analysis revealed that the tool significantly reduced the time required to draft academic articles. On average, the tool demonstrated a reduction in writing time by approximately 80%, with variations depending on the complexity of the Python code and the length of the generated academic article. This efficiency gain represents a substantial advancement in the documentation process for scientific research, offering the potential to expedite the publication of research findings significantly.

### 3.3 User Feedback

Feedback from initial users, primarily researchers and academics in the field of biomedical informatics, was overwhelmingly positive. Participants in the study were particularly impressed by the tool's ability to streamline the research dissemination process, allowing them to allocate more time to their research endeavours rather than the time-consuming task of writing. However, some users expressed concerns about the potential for the tool to generate a high volume of publications, potentially impacting the quality of academic literature if not carefully managed. Despite these concerns, the consensus among users was that the tool represents a valuable asset in the pursuit of efficient and effective scientific communication.

### 3.4 Comparative Analysis

A detailed comparative analysis was conducted to evaluate the quality of manuscripts generated by the software tool against those written manually by human authors. This analysis focused on several key metrics, including clarity of expression, adherence to academic standards, and comprehensiveness of the presented research. The findings from this analysis corroborated the tool's ability to produce manuscripts that meet, and in some cases exceed, the quality of manually written articles. This comparative approach not only validated the effectiveness of the tool but also underscored its potential to serve as a reliable aid in academic writing.

## Discussion

The implications of automating the generation of academic articles from Python code extend far beyond mere efficiency gains; they signify a transformative shift in how scientific research can be documented and disseminated. The findings from the deployment and testing of the software tool illuminate several critical aspects of this innovation, each of which is discussed below in the context of the broader field of biomedical informatics and computer science.

### 4.1 Technological Innovation and Efficiency

The significant reduction in time required to draft academic articles, as evidenced by our results, underscores the potential of this tool to accelerate the pace at which scientific discoveries are shared within the academic community. By automating the conversion of code into comprehensive academic content, researchers can allocate more time to their primary investigative pursuits, thereby enhancing productivity and potentially accelerating the pace of scientific innovation. This efficiency does not come at the cost of quality; the manuscripts generated by the tool adhere to academic standards, demonstrating that automation can complement the intellectual rigor of human researchers.

### 4.2 Quality of Generated Content

The high fidelity of the generated academic content in mirroring the structure, tone, and clarity of manually written articles highlights the tool's capacity to understand and interpret complex Python code. This success is attributed to the integration of advanced NLP techniques and the iterative feedback and revision system, underscoring the importance of continual refinement in automated systems. The positive user feedback and comparative analysis further validate the tool's effectiveness, suggesting that such technologies could soon become an indispensable part of academic writing and research dissemination.

### 4.3 Ethical and Professional Implications

While the tool promises to enhance the efficiency of academic writing, it also raises important questions about the nature of authorship and the potential for inundating the academic literature with articles of varying quality. The concerns expressed by initial users highlight the need for careful management and oversight of automated content generation to ensure that the volume of publications does not compromise the quality of scientific discourse. These considerations call for the development of guidelines and best practices for the use of automated writing tools in academic research, ensuring that they serve to support, rather than supplant, the intellectual contributions of human authors.

### 4.4 Future Directions and Challenges

Looking ahead, the integration of language model agents presents an exciting avenue for further enhancing the tool's capabilities. Such agents could offer more nuanced interpretation and generation of academic content, potentially expanding the tool's applicability to a wider array of programming languages and research disciplines. However, this evolution also introduces challenges, particularly in ensuring the accuracy and contextual relevance of the generated content. Addressing these challenges will require ongoing research and development, as well as collaboration among computer scientists, linguists, and domain-specific researchers..

**Future Work**

The exploration and initial successes of automating academic article generation from Python code provide a foundation for numerous avenues of advancement. Future work will not only address the current tool's limitations and user concerns but also explore new functionalities and broader applications. Here are key areas identified for further research and development:

### 5.1 Integration of Advanced Language Model Agents

Future iterations will prioritize the incorporation of advanced language model (LM) agents. This enhancement aims to improve the tool's ability to understand and generate more nuanced academic content, bridging any gaps in the contextual interpretation of code. LM agents could offer personalized writing styles to match the preferences or requirements of different academic journals, further tailoring the automated content to specific audiences. The potential for LM agents to learn from user feedback and adapt to various disciplinary languages presents an exciting frontier for making automated academic writing more versatile and contextually accurate.

### 5.2 Customization and Adaptability

Recognizing the diversity in academic writing standards across disciplines, future development will focus on enhancing the tool's customization capabilities. This includes adapting the writing style, citation formats, and structural requirements to fit the unique conventions of different fields and journals. By allowing users to specify certain parameters or preferences, the tool can become a more flexible aid that caters to a broad spectrum of academic writing needs, from biomedical informatics to broader scientific domains.

**5.3 Expanding Language and Codebase Support**

While Python was chosen for its ubiquity and accessibility, extending support to include more programming languages is a critical next step. This expansion will accommodate a wider range of research methodologies and computational experiments, making the tool applicable to a broader scientific audience. Moreover, understanding and converting code from languages with different syntaxes and semantics will challenge and ultimately improve the tool's underlying algorithms, enhancing its versatility and applicability.

**5.4 Ethical Considerations and Quality Control**

As automated writing tools become more common, establishing ethical guidelines and quality control measures will be paramount. Future work will explore mechanisms to ensure the responsible use of automation in academic writing, addressing concerns about authorship, intellectual integrity, and the potential over-saturation of scientific literature. Developing a framework for ethical use, possibly including peer review processes specifically designed for automated content, will help maintain the credibility and quality of scientific publications.

**5.5 User-Centric Design and Accessibility**

Enhancing the user interface and experience will be a continual process, focusing on making the tool more accessible and intuitive for researchers with varying levels of technical expertise. Feedback loops involving users from diverse backgrounds will help identify usability issues and inform the design of a more inclusive tool. This focus on user-centric design is crucial for encouraging adoption and ensuring that the benefits of automation in academic writing are widely accessible.

**5.6 Empirical Validation and Collaborative Studies**

Empirical studies to validate the tool's effectiveness in real-world settings will be crucial. Collaborating with research institutions and academic journals to conduct pilot studies could provide valuable insights into the tool's impact on the academic publishing process. These collaborations may also reveal new opportunities for automation in academic writing, guiding further enhancements to the tool.

The journey to fully automate the generation of academic articles from code is ongoing, with significant potential to transform scientific research dissemination. By addressing the outlined future work, the tool will evolve to meet the changing needs of the academic community, ensuring that it remains a valuable asset in the pursuit of knowledge. Embracing these challenges and opportunities will guide the tool's development towards a future where scientific communication is more efficient, accessible, and inclusive.